\documentclass[conference, 10pt]{IEEEtran}
\usepackage{cite}
\usepackage{amsmath,amssymb,amsfonts}
\usepackage{algorithm}
\usepackage{algpseudocode}
\usepackage{graphicx}
\usepackage{textcomp}
\usepackage{float}
\usepackage{url}

\begin{document}

\bibliographystyle{ieeetr}

\title{Vermillion: A High-Performance Scalable IoT Middleware for Smart Cities}
\author{\IEEEauthorblockN{Poorna Chandra Tejasvi\IEEEauthorrefmark{1},
Vasanth Rajaraman\IEEEauthorrefmark{2}, Arun Babu Puthuparambil\IEEEauthorrefmark{3}, Akhil Pankaj\IEEEauthorrefmark{4},
Bharadwaj Amrutur \IEEEauthorrefmark{5}}
\IEEEauthorblockA{Robert Bosch Centre for Cyber-Physical Systems,
Indian Institute of Science,\\
Bangalore, India\\
Email: \IEEEauthorrefmark{1}poornachandra@iisc.ac.in,
\IEEEauthorrefmark{2}vasanthr@iisc.ac.in,
\IEEEauthorrefmark{3}barun@iisc.ac.in,
\IEEEauthorrefmark{4}akhilpankaj@iisc.ac.in,
\IEEEauthorrefmark{5}amrutur@iisc.ac.in}}

\maketitle

\begin{abstract}
With the massive increase in the number of IoT devices being deployed in smart cities, it becomes paramount for middlewares to be able to handle very high loads and support demanding use-cases. In order to do so, middlewares must scale horizontally while providing a commensurate increase in availability and throughput. Currently, most open-source IoT middlewares do not provide out-of-the-box support for scaling horizontally. In this paper, we present ``Vermillion'', a scalable, secure and open-source IoT middleware for smart cities which provides in-built support for scaling-out. We make three contributions in this paper. Firstly, the middleware platform itself along with a formal process for data exchange between data producers and consumers. Secondly, we propose the use of hash-based federation to distribute and manage load across various message broker nodes while eliminating inter-node synchronisation overheads. Thirdly, we discuss a case study where Vermillion was deployed in a city and briefly discuss about deployment considerations using the obtained results.
\end{abstract}


\section{Introduction}
Metropolitan cities face increasing pressure on their already limited resources due to unplanned allowance of rapid urbanisation. This leads to urban sprawls and unsustainable loads on critical resources in cities \cite{tian2017impacts}. A promising solution to the aforementioned problems is the use of IoT technologies to regulate unchecked resource consumption. Although the use of IoT does not completely solve the problem, it significantly ameliorates it, thus paving the way for reasonably sustainable growth of cities \cite{su2011smart}

The use of IoT in cities comes with immense benefits such as being able to actively monitor services, resource usage, potential dangers etc. thus enabling city administrators to take prophylactic steps in preventing problems from occurring \cite{villanueva2013civitas}. Therefore, it is no surprise that the number of IoT devices will reach 18 billion by 2022 \cite{rostanski2014evaluation}. To be able to handle such rapid growth, it is very essential for IoT middlewares to technologically adapt to such demanding circumstances. There will, inevitably, be very large loads on these middlewares and they will be expected to perform with near-zero downtime while at the same time ensuring a high throughput. 

To address this problem, we have developed Vermillion\cite{vermillion}, a scalable, secure and open-source IoT middleware, built for performance.  Vermillion comprises of numerous containerised microservices all of which work together to provide a view of a single logical entity. It uses Vertx.io \cite{vertx} for the HTTPS proxy, RabbitMQ \cite{rabbit} as the message broker, PostgreSQL \cite{postgres} as the authentication database and MongoDB \cite{mongo} as the historical data store. All of these microservices have been containerised using Docker \cite{docker} and are orchestrated either using docker-compose (for a single node deployment) or using docker swarm and Ansible \cite{ansible} (for a multi-node deployment)

The primary focus of Vermillion was performance and functionality. We developed this middleware over multiple iterations, learning lessons from each, and we have finally arrived at a set of tools and practices that work well together to provide good performance results. The first tool that significantly aided in giving a performance boost was Vertx. Most traditional network libraries and frameworks spawn one thread per concurrent user. This thread dies after the user has finished their operation or is returned to the thread pool if it is a slightly more optimised implementation. Although it is a very straightforward solution to handling concurrency and load, it comes with a lot of overheads of thread and resource management. With a lot of concurrent users, there is only up to a point until which vertical scaling helps alleviate this problem. Beyond that, thread scheduling and resource management become very expensive. 

\begin{figure}[H]
\includegraphics[width=\linewidth, height=\linewidth, keepaspectratio]{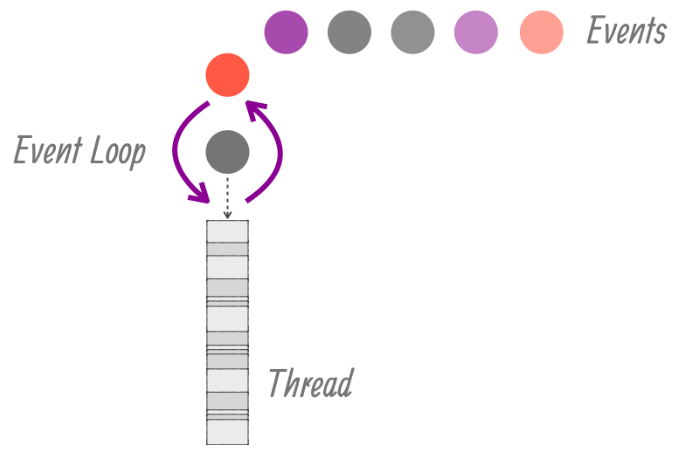}
\caption{Event loops in Vertx\cite{eventloop}}
\label{eventloop}
\end{figure}

Vertx is an event-driven and non-blocking toolkit that takes a different approach towards handling this problem.  Events are processed over event loops which are non-blocking. Each event loop has a dedicated thread associated with it. Vertx spawns two event loops per CPU core of the machine. Events get processed in a stipulated amount of time and are not allowed to block the event loop. This approach, as shown in Figure \ref{eventloop}, along with several other optimisations in the Vertx core library, which we cannot describe here due to a space constraint, has allowed it to outperform most other web servers along various measurement parameters \cite{techempower}. 

The second optimisation that we used to improve broker operation performance is by maintaining a connection pool in the HTTPS proxy. Typical RabbitMQ operations take anywhere between 14-19 TCP packets \cite{cloudamqp}. These operations include opening a connection, opening a channel, performing the operation, closing the channel and finally closing the connection. Connection pools greatly reduce connection establishment latencies for clients since they reuse connections and channels. When a client connects for the very first time, the initial overhead is incurred and the created channel is stored in memory. For every other subsequent operation, the channel is retrieved from the pool and is used for the desired client operation.

The third optimisation, especially in the multi-node deployment, is the use of a simple hash function to distribute load across many RabbitMQ nodes. We describe the hash function in detail in section \ref{hashalgo}. To compare the results we obtained from the hash-based scheme, we set up an identical cluster using RabbitMQ's internal clustering tool and we measure the throughput in both these cases.

\section{Related Work}
A variety of open-source IoT middlewares are discussed in \cite{ray2016survey} and \cite{farahzadi2018middleware} and compared extensively on various aspects like Application Development, Data Management, System Management, Heterogeneity Management, Analytics, Visualisation and so on. In both of the papers, the necessity of a middleware in an IoT environment is highlighted.  A few notable middlewares from both \cite{ray2016survey} and \cite{farahzadi2018middleware} includes ThingsSpeak, FIWARE, Xively. These middlewares however, are more focused on providing features for easy application development, visualisation and data management. 

A Service-Oriented Architecture (SOA) based Multi-Agent System (MAS) middleware \cite{zhiliang2011soa} combines the advantages of both SOA and MAS to deliver a viable performing middleware which can service 60 requests per second. The proposed architecture was able to convert a heterogeneous network into a homogeneous one system which enjoys added benefits of loosely coupled, robust, versatile, flexible and dynamic components.

According to \cite{da2018performance} and \cite{cardoso2017benchmarking} middlewares can be analysed on qualitative as well as quantitative dimensions. Qualitative aspects can include metrics such as data security, number of SDKs, number of updates available, communication protocols, support, documentation and so on. Quantitative dimensions can include response times, size of data/packet, error percentage etc. Inatel, a proprietary IoT middleware, is compared against 4 other open-source middlewares on the aforementioned aspects, out of which Fiware with Orion performed the best [4]. Similarly, FIWARE is compared with ETSI M2M [5] where FIWARE performs overall well with minimal content-length in the HTTP request and response messages. ETSI M2M suffered serious implementation flaws due to library errors and the absence of a reference implementation.

MediaSense \cite{kanter2012mediasense} is another open-source middleware which offers scalable, seamless and real-time access to sensors and actuators. The paper discusses the various requirements that a generic IoT middleware should satisfy. Proof-of-concept applications have been developed to demonstrate its capabilities in health monitoring, object-tracking, energy consumption, etc. No quantitative assessments were conducted whatsoever to support these claims. 

Performance evaluation of RabbitMQ when subjected to intense load is conducted in \cite{vandikas2014performance} and \cite{rostanski2014evaluation}. Instead of comparing the IoT-framework described in \cite{vandikas2014performance} with other middlewares, it establishes a baseline for performance on a single node system. 
Even though a clustered architecture was discussed, the performance of only a single node system on RabbitMQ was evaluated in \cite{vandikas2014performance} which is usually not the case when it comes to actual deployments. A varying number of producers and consumers were simulated, generating $10^5$ HTTPS requests, throughput (msgs/secs)  and memory consumption were recorded. The numbers showed linear scalability but were still not satisfactory.

A variety of configurations for RabbitMQ were presented in\cite{rostanski2014evaluation} to achieve High Availability and Fault-Tolerant system. A setup similar to the one mentioned in \cite{vandikas2014performance} was adopted with varying number of clients and publishers, and throughput(msgs/sec) was captured. These configurations though suffered significant performance degradation on a 3 node clustered system.

One common drawback on the systems mentioned above and in \cite{ray2016survey,farahzadi2018middleware} is that, firstly, the platforms focus on Integrated Command and Control Centres (ICCC) dashboards where the emphasis is on data acquisition and visualisation rather than on data exchange. Secondly, most of them are not completely open-source and performance numbers are not reported. So planning for a city-scale deployment becomes very difficult. Thirdly, in many platforms, the concept of standardised data exchange through REST APIs is not defined.
For middlewares where performance numbers are available \cite{zhiliang2011soa}\cite{da2018performance}\cite{cardoso2017benchmarking}\cite{vandikas2014performance}\cite{rostanski2014evaluation} they fell short on performance, scalability or both. With this paper, we try to address all of the aforementioned drawbacks and provide details on how Vermillion addresses these issues in detail.

\section{System Architecture}

We now present detailed system-level architectural details of Vermillion, both for a single-node as well as for a multi-node deployment. Vermillion comprises multiple modular components, or microservices, each of which is responsible for a specific set of functions. All of these microservices have been packaged as portable docker containers, so that setup and installation overheads are significantly reduced. The microservices have been containerised such that they intelligently wait for their dependent services to start-up before starting-up themselves whether on a single node deployment or a multi-node deployment. This approach obviates the need for microservice level orchestration, further reducing installation time.  A description of the individual microservices is as follows:

\begin{itemize}
\item \textbf{HTTPS Proxy}: This is the module that is responsible for the main application logic of Vermillion. It maintains the connection pools, computes the hash function for multi-node deployment and also round-robins between broker nodes in the cluster-mode deployment. It provides wrappers for all low-level broker operations such as publish, subscribe, bind, unbind and so on. It is designed such that it is stateless so that as many replicas can be spawned as needed without worrying about data integrity between proxy servers. It is written purely in Java using the Vertx toolkit.  

\item \textbf{Message Broker (Broker)}: We use RabbitMQ as the message broker for our middleware. RabbitMQ is a high-performance message broker written in Erlang \cite{ionescu2015analysis} that has a good throughput and scales very well\cite{rabbitisnice}. It is very customisable and provides fine-grained access-policy authoring features. For authentication and authorisation, we use an HTTP authorisation backend along with a cache plugin that keeps authorisation policies cached for quicker access.  

\item \textbf{Authentication and Authorisation Database (Auth DB)}: This is a PostgreSQL database that is used to store user IDs, hashes of API keys and permissions. The tables are designed in such a way there are no joins involved. This approach helps to reduce latencies involved in complex join operations across indices of multiple tables. This database is also used by the RabbitMQ HTTP backend to authenticate users.

\item \textbf{Broker Authorisation Backend (Auth Backend)}: RabbitMQ can be configured to use a variety of ways to authenticate and authorise users. We chose to configure it to use an HTTP backend because the logic for the complex permission systems of Vermillion could not be expressed using RabbitMQ's internal policies. Secondly, since the Auth DB was already storing authentication and authorisation information we did not want to create a split-brain situation by mirroring these policies in the broker's internal database. Hence, we chose to use an HTTP auth backend that simply connects to the Auth DB and provides a wrapper for Vermillion permission systems in the form of APIs for the broker to invoke.

\item \textbf{Historical Data Store}: We use MongoDB as the historical data store which stores data points from all connected devices. MongoDB is a NoSQL database that is simple to use, light on memory usage and has good query capabilities \cite{mongo}. We currently do not expose this data for general use to consumers. It is only meant for archival purposes.

\item \textbf{Utility Services}: This comprises of two modules: the DB connector and the unbind daemon. The DB connector is responsible for pulling out messages from all broker nodes and pushing it into the historical data store. The unbind daemon is responsible for periodically checking for expired entries in the Auth DB and deleting bindings between corresponding exchanges and queues in the broker.
\end{itemize}

\subsection{Deployment Options}
\begin{figure}
\includegraphics[width=\linewidth, height=\linewidth, keepaspectratio]{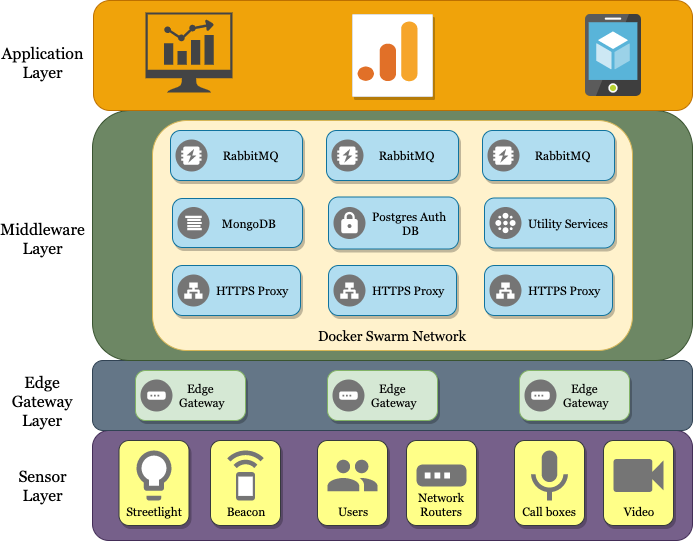}
\caption{Multi-Node Architecture}
\label{multinode}
\end{figure}

Vermillion can be deployed in three modes - a single-node deployment, a multi-node clustered deployment or a multi-node federated deployment which uses the hash-based federation algorithm. Fig. \ref{multinode} illustrates the multi-node deployment of Vermillion. 

In the single-node deployment, all docker containers run on the same machine. While this is a quick way to get Vermillion running, it is only recommended for small cities and development environments. While this mode of deployment provides a good throughput, it may not be enough for a large city or for a city with a very high number of sensors.

In the clustered deployment mode, all broker nodes are replicas of each other i.e. exchanges and bindings exist on all nodes of the cluster. Queues reside on the node on which they were created. RabbitMQ automatically routes consume requests to the respective node on which the queues reside. We do not use queue mirroring since enabling this option causes a significant decrease in throughput. Clustering option can be used to achieve fault tolerance by configuring mirrored queues at the expense of throughput.

Vermillion can also be configured to run in a federated mode where a simple hash function is used to load balance between the broker nodes. Using this technique, each broker node can run independently, while the orchestration logic can reside in the HTTPS proxy thus eliminating the need for broker level data synchronisation and state-awareness. 

\section{The Federation Algorithm} \label{hashalgo}
The federation algorithm is a simple modular hash function that uses the first character of a username to distribute load among various message brokers. The algorithm obtains a numerical equivalent of the first character of the username to compute the bucket number of the broker. This is done using the ASCII \cite{ascii} representation of the letter. Once the bucket number has been obtained, the corresponding URL is constructed by appending the bucket number to the base URL ``rabbit" and is returned to the calling function which establishes a connection with the respective broker. The pseudo-code is shown in Algorithm \ref{hash} 

Line \ref{mainline} in the algorithm denotes the main logic of the hash. For each $k\in\{1,\ldots,26\}$ and for each nodes $=2,\;3,\ldots$, we define  

\begin{equation}
bucket = (k \; \% \; nodes) \; + \; 1.
\end{equation} 

It is easy to prove that the computed bucket number always lies in $[1,nodes]$ Since $k \; \% \; nodes$ always lies in $[0,nodes-1]$, it follows that $(k \; \% \; nodes) +1$ lies in $[1,nodes]$. Hence, $bucket \in [1,nodes]$.

The hash algorithm used is a really simple one that neither uniformly distributes the load nor is consistent when nodes are added or removed from the deployment. However, we wanted to underscore the benefits of using a hash function, and how it helps in improving the overall throughput, rather than on the performance and consistency of the algorithm itself. Vermillion has been designed such that any hash function can easily be plugged in and used.

\begin{algorithm}[H]
\caption{Hash based Federation}\label{hash}
\begin{algorithmic}[1]
\Procedure{computehash}{$username, nodes$}
\State $\textit{firstchar} \gets \text{lowercase(}\textit{username}\text{[0])}$
\State $k \gets \text{ASCII value of } \textit{firstchar}$
\State $k \gets k - 96$ \Comment{Map letters to integers, a=1, b=2 etc}
\State $bucket \gets (k  \; \% \; nodes) \; + \; 1$ \Comment{$bucket \in [1,nodes]$} \label{mainline}
\State \Return \textit{bucket}
\EndProcedure
\end{algorithmic}
\end{algorithm}

\section{Data Exchange Flow}
We now turn our attention to the mechanisms that Vermillion provides to ease the process of data exchange between data providers and data consumers in a smart city. Most IoT middlewares, as stated earlier, focus on the Integrated Command and Control Centre (ICCC) dashboard for city administrators rather than creating an ecosystem for data to be exchanged between parties. When cities do expose data, they are in the form of static files. This kind of set up precludes private parties from participating and monetising any insights they generate out of the public datasets.

Vermillion, not only allows for static files to be hosted, but also for real-time data to be shared between providers and consumers. Furthermore, the data providers can choose who they share the data with, and how much of data they have access to; thus allowing for data to be monetised. The following subsections will detail this formal data exchange procedure.

Vermillion has four categories to which a user can belong. 

\begin{itemize}
\item \textit{Administrator} - A user who is an administrator is of the Vermillion server.
\item \textit{Provider} - A provider is a user who owns certain datasets and/or devices.
\item \textit{Entity} - An entity is any sensor, pseudo-sensor or application that is owned by a provider
\item \textit{Subscriber} - A subscriber is a special type of entity which mostly subscribes to data. A subscriber also must be owned by a provider.
\end{itemize}

\subsection{Resource Registration}

\begin{figure}[H]
\includegraphics[width=\linewidth, height=\linewidth, keepaspectratio]{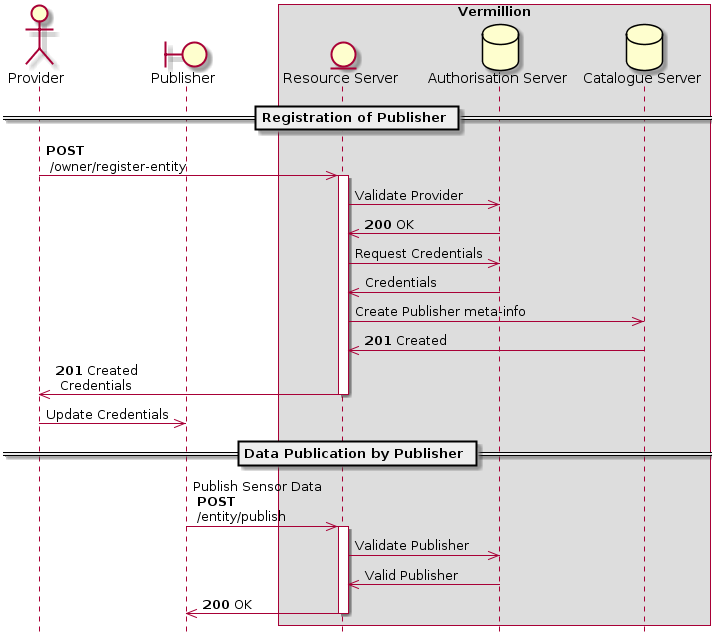}
\caption{Registration and Publish Workflow}
\label{registration}
\end{figure}

A Provider is validated and registered with Vermillion by the administrator. Upon registration, a provider API-key is generated for the given provider-id. This API key forms the basis for identity establishment of the provider. The generated API key is hashed and stored in the PostgreSQL database. With this, the provider can register entities (publishers and subscribers) with Vermillion. The work-flow for registration of a publisher and data publication is shown in the Figure \ref{registration}. 

\begin{itemize}
\item To register a publisher, the provider calls the ‘/owner/register-entity’ API with a JSON structured catalogue-item which consists of meta-information about the device e.g., its location, the type of data it publishes and so on. Vermillion validates this and accepts the request for registration by the provider by registering the entity and responding with an entity API-key.

\item To publish data, the publisher calls the ‘/entity/publish’ API with the data structured as per the data-model provided in the catalogue-item. Vermillion validates the request and publishes it to the corresponding exchange in the broker.
\end{itemize}

\subsection{Data Access Request Workflow}

\begin{figure}
\includegraphics[width=\linewidth, height=\linewidth, keepaspectratio]{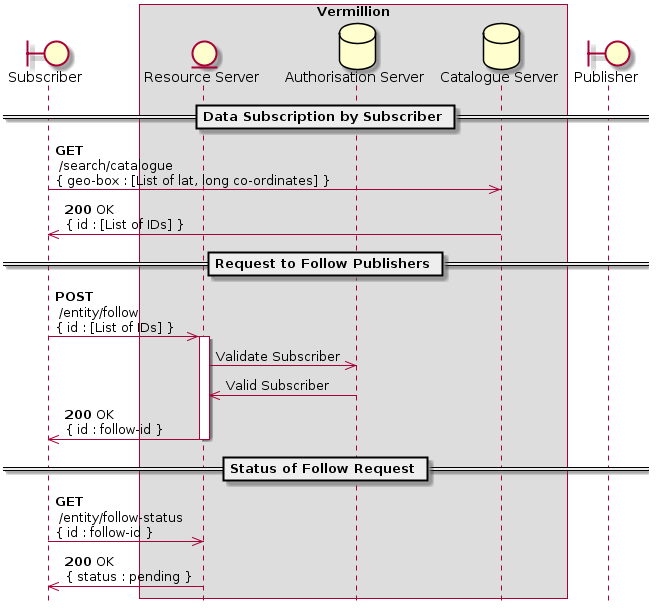}
\caption{Data Access Request Workflow}
\label{follow}
\end{figure}

As stated earlier, a subscriber is an entity which is interested in datasets belonging to various providers. A subscriber can be an application, a pseudo-sensor, or itself be a publisher. The information about all the the datsets are available in the catalogue. (at the /cat endpoint) This information can be filtered using various search parameters like geo-bounds, providers, device type and so on. Once the subscriber has a list of entities that they are interested in, they can invoke the ‘/entity/follow’ API. This API is meant for expressing interest to get access to datasets. Vermillion responds with a follow-id, which can be used by the subscriber to track the status of the request using the ‘/entity/follow-status’ API. This is illustrated in Figure \ref{follow}

\subsection{Notification of Follow request}

\begin{figure}
\includegraphics[width=\linewidth, height=\linewidth, keepaspectratio]{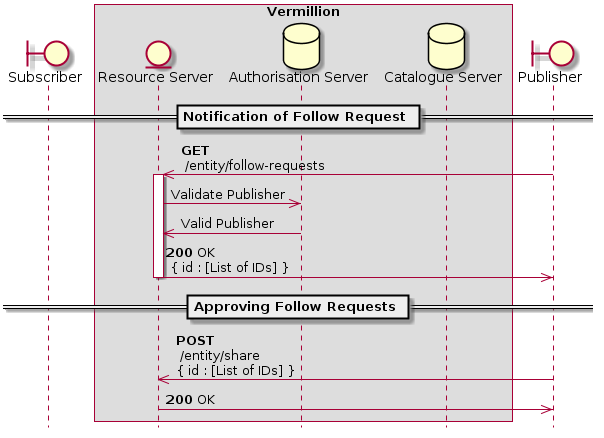}
\caption{Notification of Follow request}
\label{notify}
\end{figure}

A publisher check for received follow requests using the ‘/entity/follow-requests’ API.  The API responds with a list of subscriber IDs along with a list of entities requested by the respective subscribers. Each follow request can be independently reviewed by the provider and can be approved or rejected. A provider can approve follow requests using the ‘/entity/share’ API and  can reject a follow request using the ‘/entity/reject-follow’ API. Figure \ref{notify} illustrates this workflow

\subsection{Subscribe for data}

A subscriber who has submitted follow requests can track the status of the requests using the ‘/entity/follow-status’ API by providing the appropriate follow-id. If the follow request has been approved, the subscriber can "bind" with the list of approved entities using the ‘/entity/bind’ API. A ``bind" operation creates a binding between the entity's exchange and the subscriber's queue. This essentially instructs the broker to relay all messages of a certain type received by the entity's exchange, to the subscriber's queue. Once bound, the subscriber can start consuming the data using the '/entity/subscribe' API.

\begin{figure}
\includegraphics[width=\linewidth, height=\linewidth, keepaspectratio]{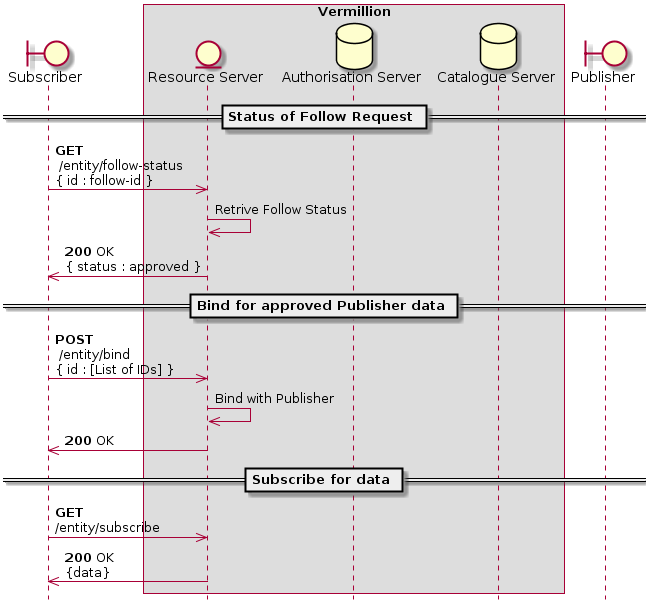}
\caption{Subscribe for data}
\label{subscribe}
\end{figure}

The bind operation is quite straightforward when it comes to dealing with a single broker node, or perhaps even a clustered RabbitMQ setup. In a clustered setup, all peers in the cluster know exactly where a particular queue resides, and routeing messages between nodes becomes easy. However, when it comes to the federated setup, some intervention is needed. All the nodes operate independently of each other. Hence, there needs to be a mechanism to allow for this routeing to happen between broker nodes. We achieve this by using the RabbitMQ shovel plugin. The shovel plugin can be used to move messages from a source on one broker node to a destination in another broker node. Vermillion keeps a track of these shovels and updates or deleted them as and when necessary. Figure  \ref{subscribe} illustrates the above mechanism.

\section{Results}

\begin{figure}
\includegraphics[width=\linewidth, height=\linewidth, keepaspectratio]{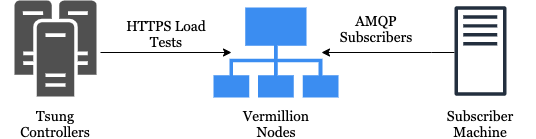}
\caption{Test Setup}
\label{testsetup}
\end{figure}

We now present preliminary test results that were performed to evaluate the throughput of Vermillion. The tests were performed on a single node deployment as well as on a multi-node deployment. The test setup consisted of three primary groups of nodes as shown in Fig. \ref{testsetup}: the Vermillion nodes, the test machines (called the controllers), and the subscriber machine. In order to load test Vermillion, we used a tool called Tsung \cite{tsung}. Tsung is a high-performance benchmarking framework built to test large-scale applications. All the nodes used for testing were high-frequency, compute optimised nodes with 16GB of RAM and 4 vCPU cores. 

\begin{table*}
\centering
\begin{tabular}{||c c c c c||} 
 \hline
 Configuration & Broker Nodes &  HTTPS Proxy Nodes & Total Nodes & Description\\ [0.5ex] 
 \hline\hline
 1 & - & - & 1 & Single Node test \\
 \hline 
 2 & 2 & 2 & 4 & Two brokers with two proxies \\
 \hline
 3 & 4 & 4 & 8 & Four brokers with four proxies\\
 \hline
 4 & 6 & 6 & 12 & Six brokers with six proxies \\
 \hline
\end{tabular}
\vspace{1em}
\caption{Test Configurations}
\label{configurations}
\end{table*}

The tests were carried out on all three deployment modes of Vermillion, namely, single-node deployment, multi-node federated deployment and multi-node clustered deployment. Table \ref{configurations} shows the configurations that were used to stress test the middleware. For each configuration, we also varied the number of producers from  $10^0$ to $10^3$, and each producer to published $10^4$ messages at a time. We did not go any further, as even for $10^4$ producers, Tsung was not able to spawn all users and manage the load. So we capped the number of producers at $10^3$. Since each producer was configured to publish $10^4$ messages, the total number of messages being published to Vermillion for a given configuration at a time, ranged from $10^4$ to $10^7$  messages\\

\begin{table}[H]
\centering
\begin{tabular}{||c c||} 
 \hline
 Number of Producers & Number of Tsung Controllers \\ [0.5ex] 
 \hline\hline
 1 & 1 \\
 \hline 
 10 & 2 \\
 \hline
 100 & 4 \\
 \hline
 1000 & 6 \\
 \hline
\end{tabular}
\vspace{1em}
\caption{Tsung Controllers per Producer Configuration}
\label{controllers}
\end{table}

Table \ref{controllers} shows the number of Tsung controllers used per producer configuration of the test. All test results conducted were over HTTPS. The throughput numbers would have been much higher with HTTP. However, we deliberately incurred this additional overhead of using TLS over HTTP, since in an actual deployment, HTTPS will likely be used over HTTP.

Furthermore, each test configuration was divided into two sub-tests. The first test measured the throughput of the system in an idealistic situation, where the broker was idle, there were no consumers, no messages in any of the queues and no running shovels. The second test attempted to model a more realistic scenario. The broker was busy catering to multiple subscribers, there were messages in many of the queues and some shovels were running. The number of consumers introduced in the second test was equal to the number of producers used for that test. E.g., if $1000$ simultaneous producers were being used in a certain test configuration, then the second test modelling a realistic load, had $1000$ simultaneously running consumers.

\begin{figure}[H]
 \includegraphics[width=\linewidth, height=\linewidth, keepaspectratio]{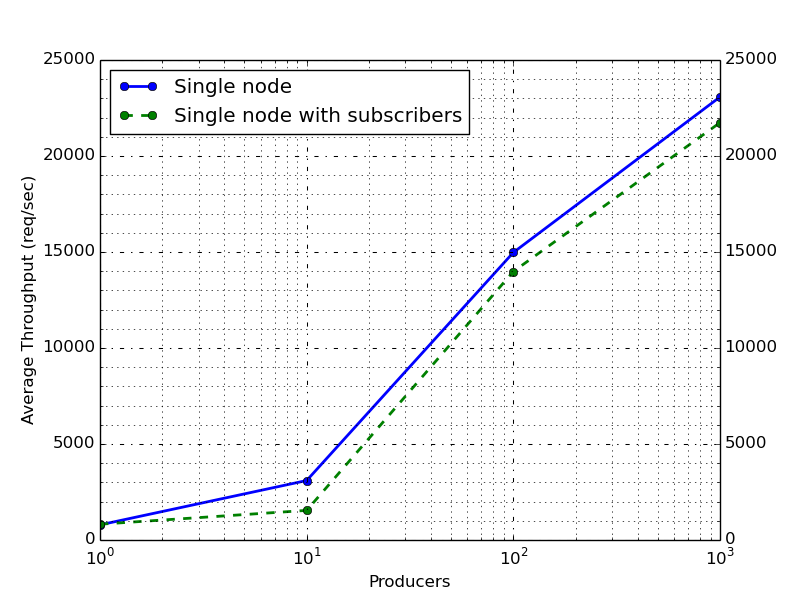}
\caption{Performance of a Single Node deployment}
\label{singlenode}
\end{figure}

Figure \ref{singlenode} shows the obtained results for a single node deployment. The maximum throughput that we obtained with this was around $23,000$ HTTPS requests per second, using $1000$ producers and no subscribers. There are a few interesting trends that we immediately observe. Firstly, the throughput increases as the number of producers are increased. This is because the HTTPS proxy is capable of handling many concurrent connections. Hence, until a point is reached where increasing the producers or the number of messages does not produce any significant impact on the average throughput, the perceived average throughput is not at its highest limit of the configuration yet. Secondly, we observe that the throughput slightly drops when subscribers are introduced. This can be attributed to the fact that there is an overhead involved in keeping long-lived connections and channels open for subscribers. We also observe that for the first test with $1$ producer and $1$ subscriber, the throughput is slightly better than when no subscribers are introduced. This could be because the number of subscribers is simply not high enough to make an impact on the overall throughput. Nevertheless, the overhead involved in keeping the connections open and their impact on the throughput is evident when the number of producers is higher. The average drop in throughput when subscribers are introduced for this configuration is about $14.6 \%$

\begin{figure}
 \includegraphics[width=\linewidth, height=\linewidth, keepaspectratio]{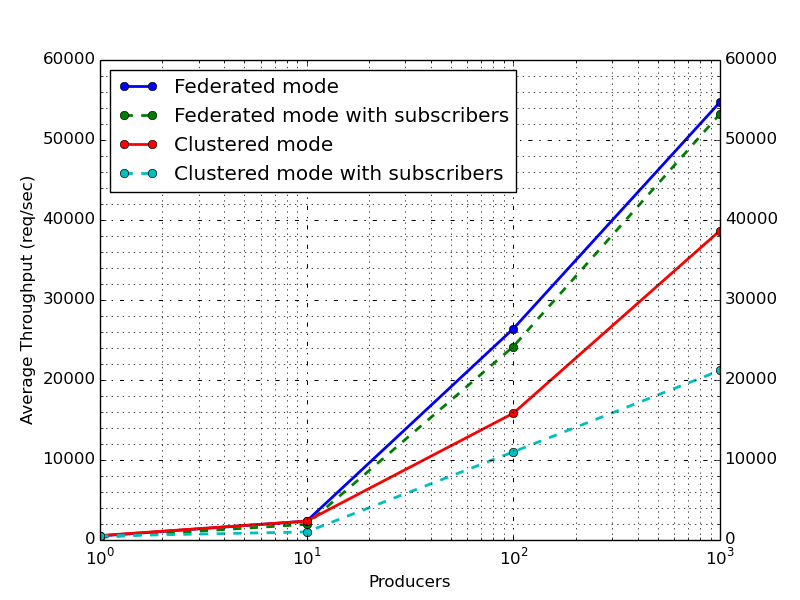}
\caption{Performance of a two-node deployment}
\label{twonodes}
\end{figure}

For two, four and six nodes, the results obtained for each configuration were compared with the throughput obtained when the respective number of RabbitMQ nodes were clustered internally. Figure \ref{twonodes} shows the results when two broker nodes and two HTTPS proxy were used. At the outset, we notice that the highest average throughput under realistic conditions for the federated mode is around $53,256$ requests per second against $21,189$ requests per second in the clustered mode. The throughput of the federated mode with subscribers is around $151\%$ better than the clustered mode. The average throughput drop in the federated mode is around $16\%$ against $36\%$ drop in the clustered setup.  

\begin{figure}
 \includegraphics[width=\linewidth, height=\linewidth, keepaspectratio]{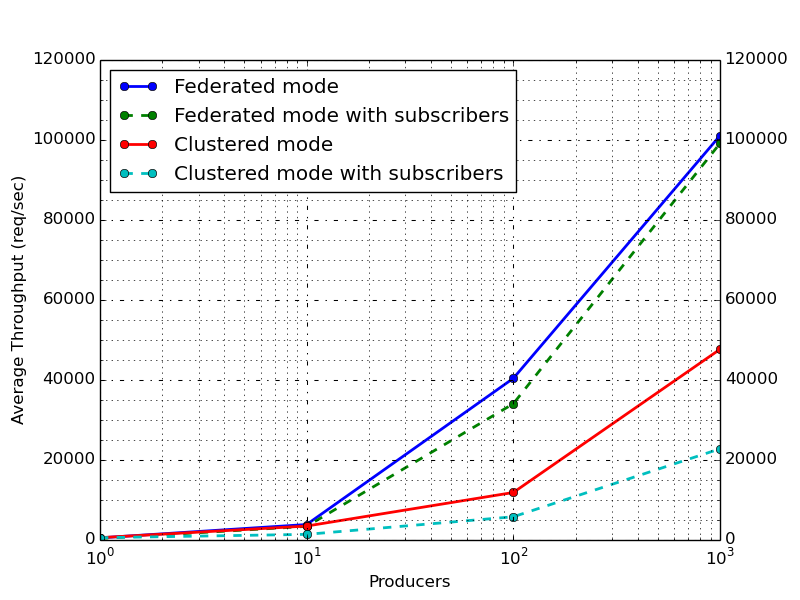}
\caption{Performance of a four-node deployment}
\label{fournodes}
\end{figure}

Figure \ref{fournodes} shows the comparison between the two modes when four broker nodes and four HTTPS nodes are used. In this particular case, the difference in the performance of the two modes is much more accentuated. The highest average throughput under realistic conditions for the federated setup is around $99,162$ requests per second against $22792.25$ requests per second in the clustered setup, which is a $335\%$ improvement in the federated mode. The average throughput drop when subscribers are introduced is around $15\%$ in the federated mode against $41\%$ in the clustered setup.  

\begin{figure}
 \includegraphics[width=\linewidth, height=\linewidth, keepaspectratio]{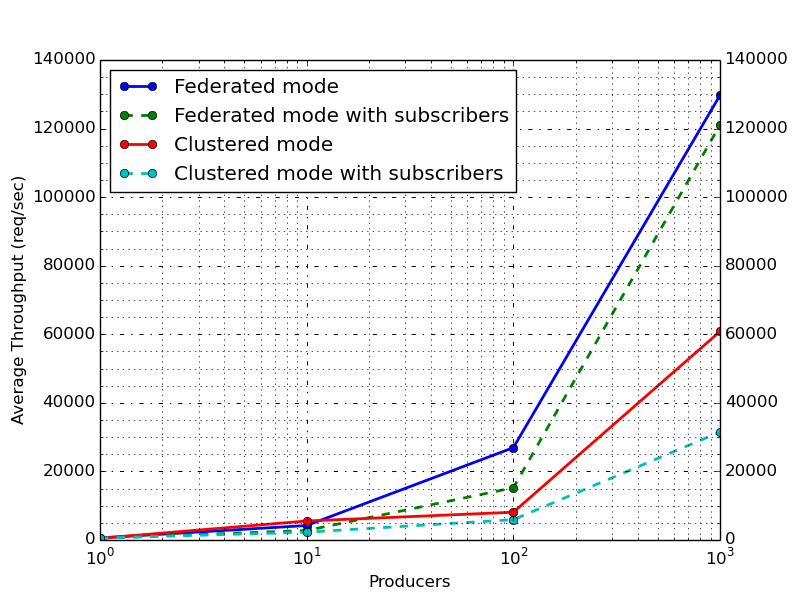}
\caption{Performance of a six-node deployments}
\label{sixnodes}
\end{figure}

When we look at the results of using six broker nodes and six HTTPS nodes, as shown in Figure \ref{sixnodes}, we notice that the throughput when $100$ producers are used, for both federated and clustered set up, is much lower than the results of their respective previous configurations. We also notice that the throughput of their next subsequent tests (when $1000$ producers are used) is much higher than the throughput obtained for their respective previous configurations. So, when using $4$ nodes and $100$ producers gave a throughput of $40,417$ requests per second in the federated setup, using six nodes and $100$ producers gave $26,933$ requests per second. This behaviour also occurs in the clustered setup. This anomaly arises because there are $12$ available nodes but only $100$ producers. Many producers finish their publication well before new producers are spawned, thus reducing the total number of concurrent publications. This causes the overall perceived throughput by Tsung to reduce.  

We notice that the highest average throughput under realistic conditions is $121,058$ requests per second for the federated mode against $31,503$ requests per second for the clustered setup, which is around a $284\%$ improvement in the federated mode. The average throughput drop when subscribers are introduced is around $25\%$ in the federated mode, and around $31\%$ in the clustered setup.

The federated mode consistently outperforms the clustered setup since there are no inter-node synchronisation overheads involved. The routing logic is embedded entirely in the HTTPS proxy. In the clustered setup, there is constant chatter between the nodes to synchronise their states, data, information about connections, channels and so on. Also, all the peers in a clustered setup need to maintain a routeing table that specifies the node to contact in case a request which cannot be served by that particular node is received. For e.g., if node-1 receives a request to subscribe for a queue residing in node-2.  

\begin{figure}
\includegraphics[width=\linewidth, height=\linewidth, keepaspectratio]{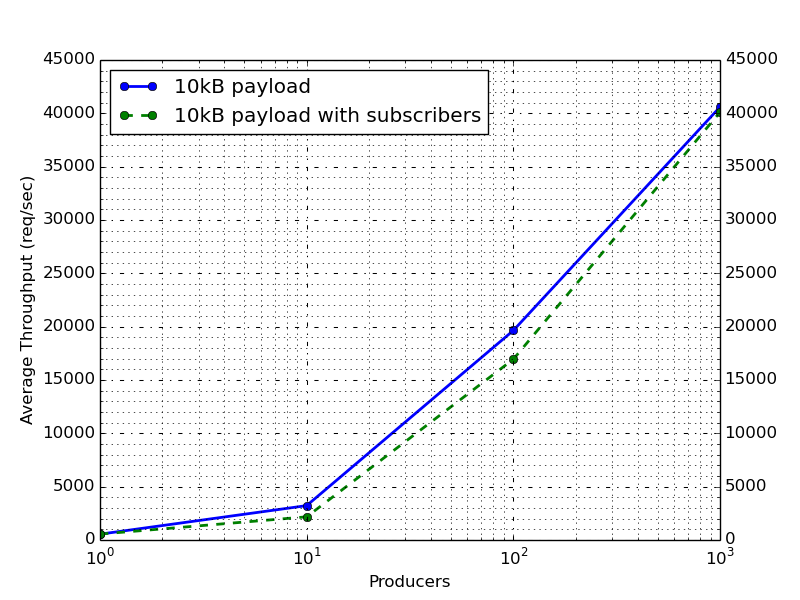}
\caption{Performance of a two-node deployment when the payload is 10kB}
\label{10kb}
\end{figure}

The payload used for the above tests was a sample reading from a flood sensor deployed in Pune, Maharashtra, India. This payload has a message size of around$220$ bytes. In many cases, sensors deployed in a smart city may not directly publish to their middlewares. They first publish to their edge gateways, which, in turn, publish to the middlewares. In such a scenario, a gateway may not publish messages as and when it receives the readings from a sensor. It may accumulate some readings over time and then publish at a certain stipulated time, thus causing the message size to increase. To simulate such a situation, we deployed the $4$ broker node, $4$ HTTPS node federated configuration and tested the setup using a $10$ kB payload. Figure \ref{10kb} shows that the average throughput drop when subscribers are introduced is around $11.5\%$. The highest average throughput obtained under realistic circumstances is $40,104$ requests per second. This throughput figure is much lesser than $99,162$ requests per second for a $220$ byte payload since the message size is much larger.

\begin{figure*}[!h]
\includegraphics[width=\linewidth, height=\linewidth, keepaspectratio]{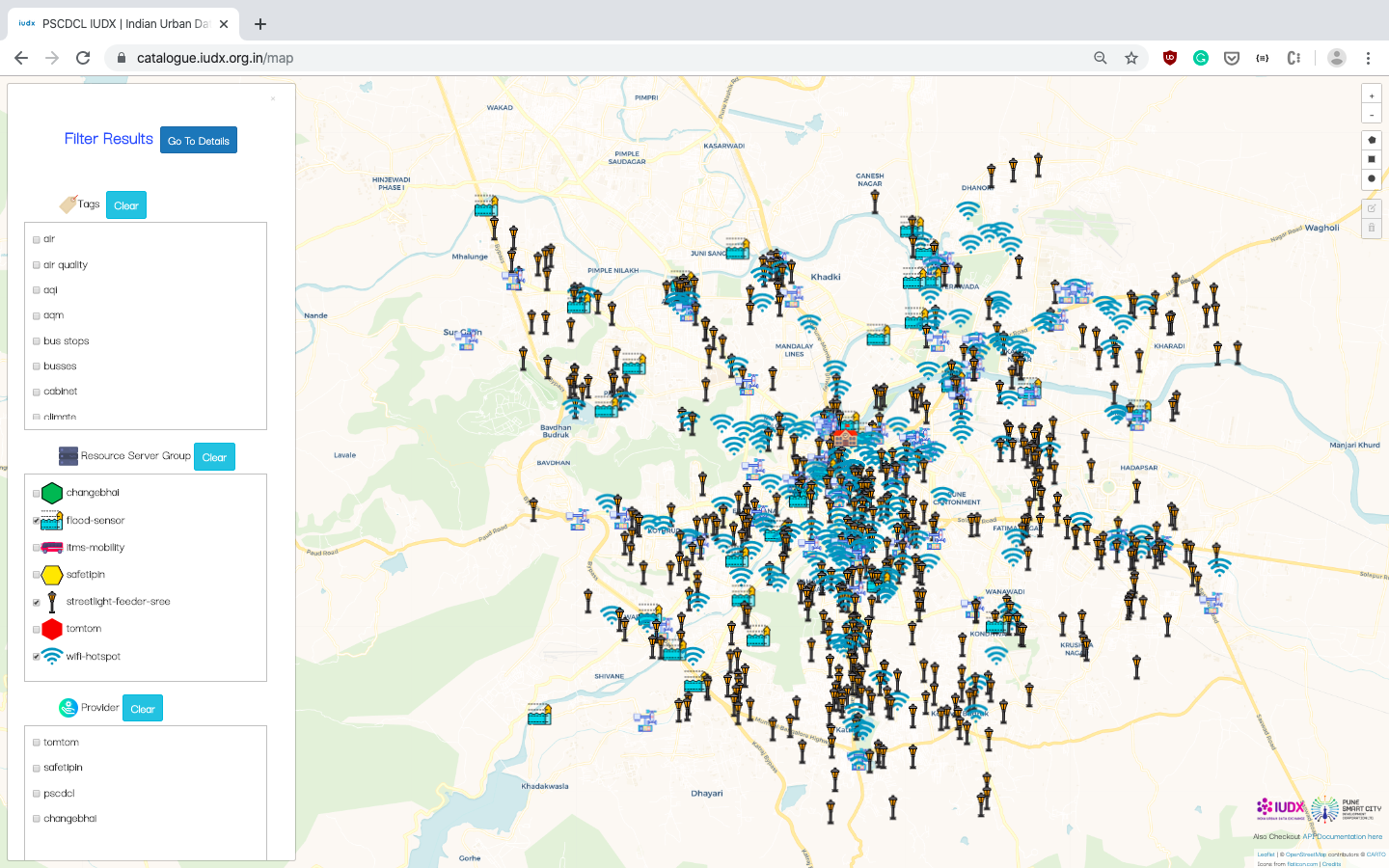}
\caption{Deployment of Vermillion in Pune Smart City serving 2200 sensors ({https://catalogue.iudx.org.in}).}
\label{map}
\end{figure*}

\section{Case Study and Deployment Considerations}
Vermillion has been deployed in the city of Pune, Maharashtra, India. It is a part of the  the Indian Urban Data Exchange \cite{iudx} deployment to provide access to smart city data in an open and secure fashion. This deployment has about $2,200$ sensors, each of them publishing at different intervals. We have set up a single node instance of Vermillion and it has been running smoothly for over two months at the time of writing this paper. The data from Vermillion can be accessed at https://catalogue.iudx.org.in. Figure \ref{map} shows a map interface of the sensors deployed in the Pune Smart City.

The desirable properties needed from a middleware might vary across different cities. So, it becomes very difficult to make a general recommendation that can suit all cities. The results obtained can be used to decide on the deployment strategy needed for a city. E.g. the city of Barcelona, Spain has around $1800$ sensors publishing data at varied periods\cite{barcelona}. For such a kind of deployment, in its current state, a single node deployment with additional customisations should suffice. On the other hand, the city of Santander, Spain has over $12,000$ sensors deployed\cite{esri}. For such a kind of setup, the sensor publish frequency must be taken into account while deploying Vermillion. Typically, the $4$ broker node, $4$ HTTPS node deployment should suffice, but this decision needs to be arrived at only after consideration of a lot of factors, including future sensor deployment plans by city administrators. 

\section{Conclusion}
In this paper, we have introduced Vermillion, a high-performance, scalable IoT middleware for smart cities. We have described the individual sub-systems of this middleware and proposed a formal procedure for data exchange between data producers and consumers in a smart city. We have also proposed the use of a hash function to federate RabbitMQ nodes and distribute load across these nodes using the hash algorithm. Furthermore, we have shown that the performance of this federated mode deployment against the internal cluster setup of RabbitMQ is at least $151\%$ better than using the internal clustering of RabbitMQ. We have also shown that Vermillion's throughput increases with increasing availability unlike the case with internal clustering of RabbitMQ. Finally we discussed a case study where Vermillion was successfully deployed in the city of Pune, Maharashtra, India.

\section{Future Scope}
As stated earlier, Vermillion can be customised and more features can be plugged in as necessary. However, there are a few features internal to Vermillion that can be improved. Firstly, a good hash function needs to be chosen so that the load is uniformly distributed among the various message brokers. The hash function also needs to be consistent so that if nodes are added or deleted from a deployment, it should not affect the functioning of Vermillion. Secondly, proxies for other low-level protocols like AMQP, MQTT, STOMP, CoAP and so on must be developed so that users and devices need not communicate only over HTTPS. Thirdly, the overall availability of Vermillion can be improved if each node in a bucket is clustered, using the RabbitMQ internal clustering, with one other node, and the Docker Swarm service round robins between these two nodes when usernames (or any other parameters) resolve to that particular bucket. E.g., if there are two buckets with one node each, these two nodes can internally cluster with one other node, thus allowing for more nodes per bucket. Docker Swarm can round-robin between the two nodes per bucket. Finally, Vermillion can allow for proper access mechanisms and retention policies for archived data.

\section{Acknowledgements}
We would like to thank the Pune Smart City Development Corporation for their support and the Ministry of Housing and Urban Affairs, Government of India for funding this project. 


\bibliography{bibliography} 

\begin{thebibliography}{10}

\bibitem{tian2017impacts}
L.~Tian, B.~Ge, and Y.~Li, ``Impacts of state-led and bottom-up urbanization on
  land use change in the peri-urban areas of shanghai: Planned growth or
  uncontrolled sprawl?,'' {\em Cities}, vol.~60, pp.~476--486, 2017.

\bibitem{su2011smart}
K.~Su, J.~Li, and H.~Fu, ``Smart city and the applications,'' in {\em 2011
  international conference on electronics, communications and control (ICECC)},
  pp.~1028--1031, IEEE, 2011.

\bibitem{villanueva2013civitas}
F.~J. Villanueva, M.~J. Santofimia, D.~Villa, J.~Barba, and J.~C. Lopez,
  ``Civitas: The smart city middleware, from sensors to big data,'' in {\em
  2013 Seventh International Conference on Innovative Mobile and Internet
  Services in Ubiquitous Computing}, pp.~445--450, IEEE, 2013.

\bibitem{rostanski2014evaluation}
M.~Rostanski, K.~Grochla, and A.~Seman, ``Evaluation of highly available and
  fault-tolerant middleware clustered architectures using rabbitmq,'' in {\em
  2014 federated conference on computer science and information systems},
  pp.~879--884, IEEE, 2014.

\bibitem{vermillion}
``{V}ermillion: {A} {H}igh {P}erformance {S}calable {IoT} {M}iddleware for
  {S}mart {C}ities.''
\newblock \url{https://github.com/rbccps-iisc/iudx-resource-server}.

\bibitem{vertx}
``Eclipse {V}ertx: {A} {T}oolkit for building reactive applications on the
  {JVM}.''
\newblock \url{https://vertx.io}.

\bibitem{rabbit}
``Rabbitmq- {M}essaging that just works.''
\newblock \url{https://www.rabbitmq.com}.

\bibitem{postgres}
``{PostgreSQL}: The world's most advanced open source database.''
\newblock \url{https://www.postgresql.org}.

\bibitem{mongo}
``Mongodb: The database for modern applications.''
\newblock \url{https://www.mongodb.com/}.

\bibitem{docker}
``Docker: {A} {C}ontainerisation tool, note={\url{https://www.docker.com/}}.''

\bibitem{ansible}
``Ansible: {S}imple {IT} {A}utomation and {O}rchestration.''
\newblock \url{https://www.ansible.com}.

\bibitem{eventloop}
``A gentle guide to asynchronous programming with {E}clipse {V}ert.x for {J}ava
  developers.''
\newblock \url{https://vertx.io/docs/guide-for-java-devs/}.

\bibitem{techempower}
``{TechEmpower} - {W}eb {F}ramework {B}enchmarks.''
\newblock \url{https://www.techempower.com/benchmarks/}.

\bibitem{cloudamqp}
``13 {C}ommon {RabbitMQ} {M}istakes and {H}ow to {A}void {T}hem.''
\newblock
  \url{https://www.cloudamqp.com/blog/2018-01-19-part4-rabbitmq-13-common-errors.html}.

\bibitem{ray2016survey}
P.~P. Ray, ``{A} {S}urvey of {I}o{T} {C}loud {P}latforms,'' {\em Future
  Computing and Informatics Journal}, vol.~1, no.~1-2, pp.~35--46, 2016.

\bibitem{farahzadi2018middleware}
A.~Farahzadi, P.~Shams, J.~Rezazadeh, and R.~Farahbakhsh, ``Middleware
  technologies for cloud of things: a survey,'' {\em Digital Communications and
  Networks}, vol.~4, no.~3, pp.~176--188, 2018.

\bibitem{zhiliang2011soa}
W.~Zhiliang, Y.~Yi, W.~Lu, and W.~Wei, ``A soa based iot communication
  middleware,'' in {\em 2011 International Conference on Mechatronic Science,
  Electric Engineering and Computer (MEC)}, pp.~2555--2558, IEEE, 2011.

\bibitem{da2018performance}
M.~A. da~Cruz, J.~J. Rodrigues, A.~K. Sangaiah, J.~Al-Muhtadi, and V.~Korotaev,
  ``Performance evaluation of iot middleware,'' {\em Journal of Network and
  Computer Applications}, vol.~109, pp.~53--65, 2018.

\bibitem{cardoso2017benchmarking}
J.~Cardoso, C.~Pereira, A.~Aguiar, and R.~Morla, ``Benchmarking iot middleware
  platforms,'' in {\em 2017 IEEE 18th International Symposium on A World of
  Wireless, Mobile and Multimedia Networks (WoWMoM)}, pp.~1--7, IEEE, 2017.

\bibitem{kanter2012mediasense}
T.~Kanter, S.~Forsstr{\"o}m, V.~Kardeby, J.~Walters, U.~Jennehag, and
  P.~{\"O}sterberg, ``Mediasense--an internet of things platform for scalable
  and decentralized context sharing and control,'' in {\em Proceedings of 7th
  IARIA International Conference on Digital Telecommunications (ICDT), 2012},
  Citeseer, 2012.

\bibitem{vandikas2014performance}
K.~Vandikas and V.~Tsiatsis, ``Performance evaluation of an iot platform,'' in
  {\em 2014 Eighth International Conference on Next Generation Mobile Apps,
  Services and Technologies}, pp.~141--146, IEEE, 2014.

\bibitem{ionescu2015analysis}
V.~M. Ionescu, ``The analysis of the performance of rabbitmq and activemq,'' in
  {\em 2015 14th RoEduNet International Conference-Networking in Education and
  Research (RoEduNet NER)}, pp.~132--137, IEEE, 2015.

\bibitem{rabbitisnice}
``Rabbitmq {H}its {O}ne {M}illion {M}essages {P}er {S}econd on {G}oogle
  {C}ompute {E}ngine,
  note={\url{https://content.pivotal.io/blog/rabbitmq-hits-one-million-messages-per-second-on-google-compute-engine}}.''

\bibitem{ascii}
``American {S}tandard {C}ode for {I}nformation {I}nterchange,
  note={\url{https://en.wikipedia.org/wiki/ascii}}.''

\bibitem{tsung}
``Tsung- high performance benchmark framework.''
\newblock \url{https://github.com/processone/tsung}.

\bibitem{iudx}
``{IUDX}: {T}he {I}ndian {U}rban {D}ata {E}xchange.''
\newblock \url{https://iudx.org.in}.

\bibitem{barcelona}
A.~Sinaeepourfard, J.~Garcia, X.~Masip-Bruin, E.~Mar{\'\i}n-Tordera, J.~Cirera,
  G.~Grau, and F.~Casaus, ``Estimating smart city sensors data generation,'' in
  {\em 2016 Mediterranean Ad Hoc Networking Workshop (Med-Hoc-Net)}, pp.~1--8,
  IEEE, 2016.

\bibitem{esri}
``{S}ensors for {S}mart {C}ities.''
\newblock
  \url{https://www.esri.in/esri-news/publication/vol9-issue1/articles/sensors-for-smart-cities}.

\end{thebibliography}

\end{document}